# A Framework of Reconfigurable Transducer Nodes for Smart Home Environments

Basim Hafidh, Hussein Al Osman, Haiwei Dong, and Abdulmotaleb El Saddik, *Fellow, IEEE*

*Abstract*— This paper presents a transducer network framework that supports the amalgamation of multiple transducers into single wireless nodes. This approach is aimed at decreasing energy consumption by reducing the number of wireless transceivers involved in such networks. To make wireless nodes easily reconfigurable, a plug and play mechanism is applied to enable the clustering of any number of transducers. Furthermore, an algorithm is proposed to dynamically detect added and removed transducers from a node. Lastly, an XML based protocol is devised to allow nodes to communicate a description of their layout, measured data and control information. To verify the proposed framework, multiple reconfigurable wireless nodes are used to monitor the dynamic condition of a multiple rooms during a period of 24 hours in order to emulate a smart home scenario.

## I. Introduction

Transducer (sensor and actuator) networks measure, process and respond to the physical environment and communicate their measured information among each other and with remote computing nodes. Several works have applied transducer networks in various domains [1-4]. This paper presents a transducer network framework for smart home environments applications.

In a smart home environment, two kinds of transducer networks can be involved, wired and wireless. Each kind has its merits and disadvantages. Wired transducer networks are restricted by their deployment as they require physical connections snaked through walls or laid through other less esthetically appealing methods; on the other hand, they consume less energy compared to wireless transducer networks as wired communication is much less energy demanding. Wireless transducer networks typically have distinct energy consumption patterns that correspond to their phase of operation: sensing, processing, and communicating. The energy consumption in the first and second phases is negligible compared to that of the communicating phase. Pottie *et al.*, for example, stated that the required energy for transmitting 1KB over 100m distance is about 3 joules and is almost the same as processing 1 million operations by a 100 MIPS/W processor [5]. Therefore, energy consumption is considered as one of the major challenges in wireless transducer network design [6, 7].

B. Hafidh, H. Al Osman, H. Dong and A. El Saddik are with Multimedia Computing Research Laboratory (MCRLab), School of Electrical Engineering and Computer Science, University of Ottawa, Ottawa, Ontario, Canada (e-mail: {bhafi014; halos072; hdong; elsaddik} @uottawa.ca).

Various mechanisms are proposed by researchers in order to minimize or optimize the energy consumption for wirelessly communicating sensors. These mechanisms can be divided into three ways: *software*, *hardware* and *communication*:

For *software solutions*, researchers have implemented algorithms in order to minimize the power consumption of wireless nodes. For instance, Chou *et al.* [10] presented a compression algorithm that makes use of previous readings in order to minimize communication. Other researchers introduced power-aware routing protocols to maximize the lifetime of the transducer nodes [11, 12].

In terms of *hardware solutions*, researchers have combined transducers together in one wireless node to reduce the number of nodes in the network and consequently decrease the energy consumption of the whole network. Several researchers successfully combined multiple transducers in one wireless node. Noh *et al.* [13], for example, presented a design for a wireless node that bundles four physically interconnected transducers. Similarly, Suh *et al.* [14] proposed a wireless node composed of three transducers, while supporting the option of attaching a fourth one through a connecter mechanism.

Since the *communication* phase of the wireless node consumes most of the energy as mentioned earlier, researchers such as Zhang *et al.*, and Surie *et al.* [8, 9] proposed the use of the energy-efficient ZigBee (IEEE 802.15.4) communication protocol as opposed to Wi-Fi or Bluetooth in wireless transducer networks used for home monitoring purposes.

Given the related literature presented above, in this work, we propose our own hardware solution to minimize the overall energy consumption of a transducer network. The existing hardware solutions, such as the ones presented by Zhang *et al.* [8] and Surie *et al.* [9], proposed fixed nodes that are limited by the type and number of transducers they support and the way the transducers are physically laid. Therefore, we present a flexible and general-purpose transducer network framework that supports the clustering of a large variety of transducers into single wireless nodes through a plug and play mechanism. Therefore, the resulting transducer network uses a hybrid of wired and wireless communication. Wired connections are used to link transducers in close proximity, which, ensures customizability for a wide range of smart home applications as it will be demonstrated in this work.

Our proposed solution can be combined with software and communication level enhancements for further energy optimization. In fact, to reduce wireless communication power consumption, we have used ZigBee as proposed by [8, 9].

## II. Framework Design

In this section, we provide an overview of our proposed

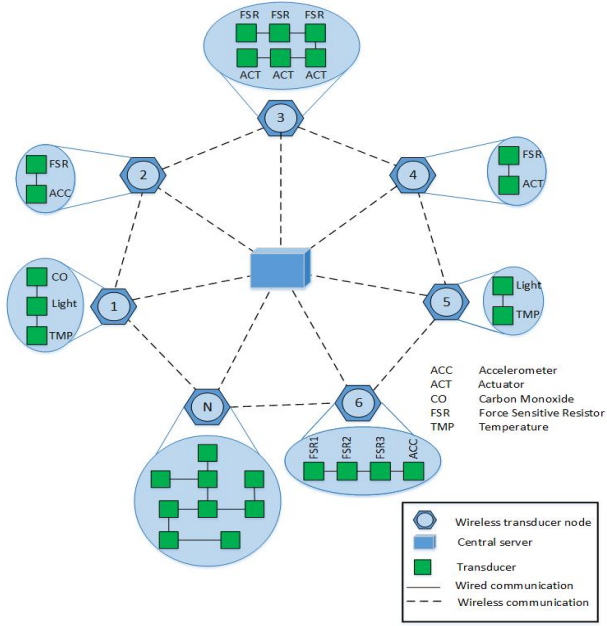

Fig. 1. Wireless nodes with reconfigurable transducers.

framework and discuss its architecture.

*A. Overview*

Our goal is to design a hybrid wired-wireless transducer network composed of several plug and play transducers. These plug and play transducers share their measured data inside the node through wires using a serial communication protocol called I$^2$C (inter-integrated circuit) and wirelessly to other nodes or central server as shown in Fig.1 through ZigBee links. The data produced by the collection of sensors on each node is hierarchically organized into an eXtensible Markup Language (XML) message that is wirelessly communicated to other nodes. Two XML messages are involved: one directed from a wireless node to other nodes or a central server which includes the node's overall information, layout description, and periodic sensory measurements (see Section 3). The other message, a control message, is received by a transducer node to activate its vibro-tactile actuators.

*B. Wireless Node Architecture*

Fig. 2 shows the architecture of the wireless node. It is composed of the following five main modules:

*1) Transducers module*

The transducers module represents a set of interconnected transducer units. Each unit is composed of a microcontroller, an analog to digital converter (ADC) or a digital to analog converter (DAC). A unique ID is assigned to each transducer (see Table I) in order to identify it. These IDs are statically pre-assigned and are permanently stored in flash memory.

*2) Communication bus*

The communication bus is responsible for providing the necessary communication between the transducers module and the data collection module. The onboard communication bus is based on I$^2$C.

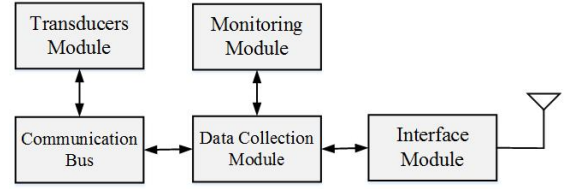

Fig. 2. Wireless node architecture.

*3) Data collection module*

This module provides the necessary space to collect all measured data that comes from the transducers module, and pass them to the interface module. It also collects the upcoming signals from the interface module, and forwards them to the transducers module in order to activate the running actuators. Moreover, this module stores the connectivity status of the node's transducers. This status includes the transducers IDs and corresponding statuses (i.e., running or not) which are forwarded to the monitoring module.

*4) Monitoring module*

This module is responsible for monitoring the connection and disconnection of transducers. When a transducer is connected to or disconnected from a node, it triggers an event by broadcasting its unique ID through the communication bus. This event is detected by the monitoring module which recognizes the ID of the transducer. The monitoring module records the change regarding the added or removed transducer and sends this information to the data collection module in order to update the corresponding transducer status.

*5) Interface module*

The interface module provides the necessary communication between the transducer node and its external environment. Five possible media of communication can be used with the proposed transducer node including: USB (RS232), Bluetooth technology, Ethernet, Wi-Fi, and ZigBee. However, the ZigBee communication technology is considered more preferable in designing sensor networks due to its low power consumption [15]. This module is responsible for exchanging the messages between the data collection module and other nodes or the central server.

## III. IMPLEMENTATION AND EVALUATION

As a proof of concept evaluation of our proposed framework, we deployed multiple wireless nodes to monitor the status of various objects and rooms in home setting. Table I lists the plug and play transducers we support in our implementation. The main board of the node is supplied by 3.3V rechargeable battery. The node's microcontroller used is Arduino pro-mini which runs at 8MHz. The Arduino software runs the data collection and monitoring modules mentioned above. This microcontroller shares the data with the wired connected transducers.

TABLE I
PLUG AND PLAY CONFIGURABLE TRANSDUCERS

| ID | Transducer | Circuit | Remarks |
|---|---|---|---|
|  | Wireless Node's Core (Mainboard) | 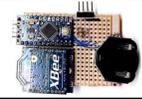 | Data collection, processing and communication |
| 1-20 | Pressure Sensor | 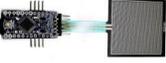 | I²C IDs 1-20 are reserved for pressure sensors (FSRs) |
| 21-40 | Vibrotactile Actuator | 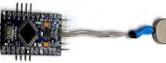 | I²C IDs 21-40 are reserved for actuators |
| 72-75 | Temperature Sensor | 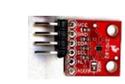 | I²C IDs 72-75 are reserved for Temperature Sensors (TMP102) |
| 83-84 | Accelerometer | 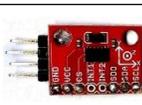 | I²C IDs 83-84 are reserved for accelerometer (ADXL345) |
| 41,57 | Light Sensor | 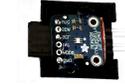 | I²C IDs 41, 57 are reserved for Light sensors (TSL2561) |
| 76-78 | CO Gas Sensor | 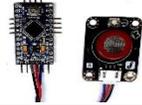 | I²C IDs 76-78 are reserved for Ambient Light sensors (TEMT6000) |
| 85-90 | Flex Sensor | 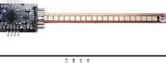 | I²C IDs 85-90 are reserved for flex sensors 4.5" |
|  | Extension | 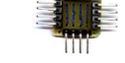 |  |

In the current implementation, the transducers cluster includes pressure (FSR, force sensitive resistor), ambient light (TEMT6000), carbon monoxide gas (MQ-7), triple axes accelerometer (ADXL 345), temperature (TMP 102), and flex sensors and vibro-tactile actuators. These plug and play transducers are implemented on top of Android pro-mini microcontrollers. Each transducer has its assigned ID number for a unique identification and is able to communicate with each node's microcontroller through the I²C serial communication protocol.

As an example, 6 wireless nodes have been located in different places in a house as shown in Fig. 3. Each wireless node has been customized depending on the measurement environment. For example, Node 1 measures the kitchen's temperature, light intensity and Carbon monoxide (CO) levels. Node 2 is placed in the refrigerator and includes pressure (placed under the egg tray as an example) and accelerometer (attached to the door) sensors. Node 3 has three sensors and three actuators mounted on a sofa's cushions. The sensors can be used to monitor sitting patterns and durations and the actuators can be used to buzz persons when they remain seated for too long. Three nodes are located in the bedroom: Node 4 with a pressure sensor and actuator mounted on a chair seat, Node 5 measures the temperature and light intensity in the room, and Node 6 located underneath the pillow with three pressure sensors and accelerometer to measure a person's movements while sleeping. These nodes and corresponding transducers are also presented in the topology of Fig.1. As we can see that the same node mainboard hardware and software are used with different plug and play transducers depending on the type of the physical property to be measured.

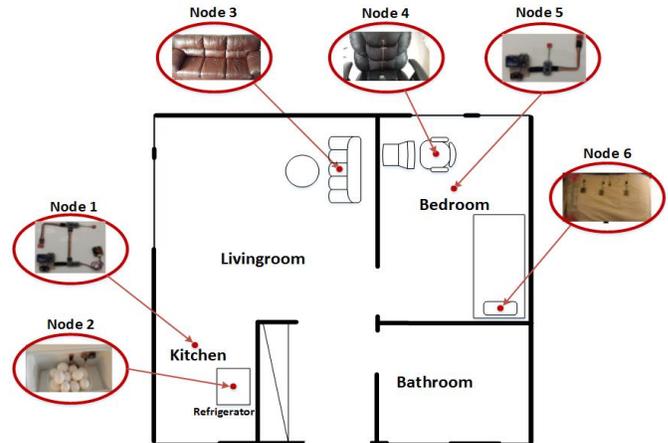

Fig. 3. A house example with 6 wireless nodes with different reconfigurable transducers.

The sampling rate of the sensors used is 1Hz except for that of the accelerometer which is 30 Hz. Each node's microcontroller sends a measurement XML message including the node's layout description (such as the communication protocol, and number and type of the transducers involved) and its running sensors' measurement (such as the timestamp and the measured data) to the other nodes or to the central server through 1mw ZigBee transceiver (digi XBee S1). The central server monitors and stores the received measurement data and post-processes this data. The server receives and stores the upcoming measured data from the nodes and prepares them for visualization to make them meaningful. It also sends control signals to the corresponding nodes in order to activate their actuators. Fig. 4 shows a visualization of the data measured by the 6 wireless nodes in the home setup previously described during a period of 24 hours. The white color represents the minimum measured value and the dark represents the maximum. For example, the first bar represents the data for the CO level in the kitchen (Node 1). This level was high from 10 am to 12 pm and from 8 pm to 9 pm (during which the kitchen was used for cooking).

In order to prove that our proposed framework consumes less energy compared to "traditional" wireless transducer networks. A traditional wireless network is a set of nodes connected through a wireless channel. Each node is composed of a single sensor, microcontroller, and wireless transceiver. Such networks have been described in numerous previous works including [8, 9, 16, 17, 18]; we have setup a dedicated experiment over a period of 24 hours. During the experiment, we ran in parallel two transducer networks: the one described above using our proposed framework and another traditional network composed of the same transducers. To measure the energy consumption, the node's supply voltage and current were measured. The current sensor (TI-INA169) was used to measure the node's drawn current. Fig. 5 shows the 24 hour energy consumption by each proposed node as compared to traditional transducer nodes that are typically used in wireless sensor networks. For example, Node 1 is compared with the combination of three traditional transducer nodes. For Node 4 (the chair node), however, there is no energy saving compared to the traditional network of nodes since Node 4 is composed of only one pressure sensor and one actuator, and the effect of the actuator was neglected as it is run for 30 seconds every 30 minutes to buzz the sitter when the pressure sensor continuously reads values greater than zero for a period of 30 minutes. We can see from the figure, the more transducers are included in the node, the more energy we save and besides the less nodes can be involved in the network. In this implementation, as the transducer node framework is standardized proposed, the designed wireless transducer was efficiently developed.

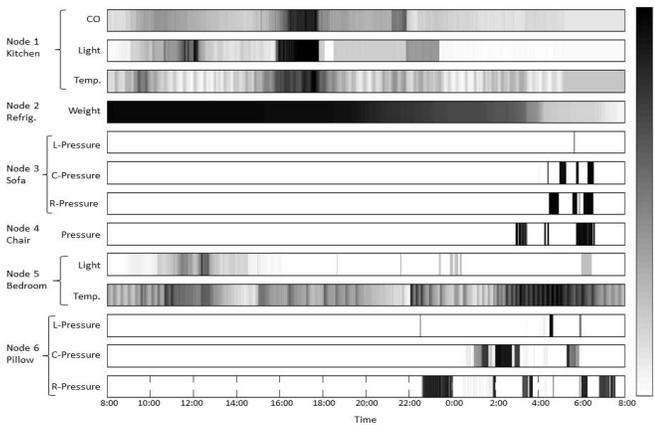

Fig. 4. Nodes sensors measurement.

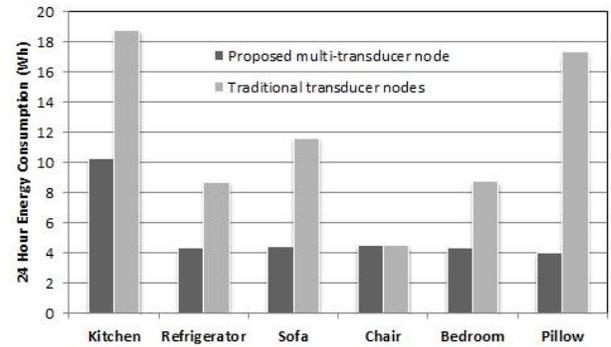

Fig. 5. Energy consumption of each proposed node compared with traditional transducer nodes.

## IV. Conclusion

In this paper, we proposed a transducer network framework composed of wireless nodes that can inter-communicate through direct links or a central server. The proposed framework was aimed at reducing the energy consumption of transducer networks through clustering. A plug and play mechanism was conceived to cluster any number of transducers together in several possible configurations. The wireless node dynamically recognized new added/removed transducers. A prototype wireless node core was implemented with several plug and play transducers. A proof of concept evaluation was done by placing different wireless nodes inside a house. Each node was composed of different transducers depending on the measured environment or object. A visualization of the data collected was shown by the various nodes dispersed throughout the home setup.

We also evaluated the energy consumption of the proposed transducer network and compared it to traditional wireless transducer networks where each node is dedicated for a single transducer. It is shown that our proposed transducer network consumes significantly less energy. For our home setup, our proposed method consumes 46% energy compared to traditional approach.